\documentclass[twocolumn,twoside,slac_two]{revtex4}
\usepackage{graphicx}
\usepackage{fancyhdr}
\def\spose#1{\hbox to 0pt{#1\hss}}
\def\lta{\mathrel{\spose{\lower 3pt\hbox{$\mathchar"218$}}
  \raise 2.0pt\hbox{$\mathchar"13C$}}}
\def\gta{\mathrel{\spose{\lower 3pt\hbox{$\mathchar"218$}}
  \raise 2.0pt\hbox{$\mathchar"13E$}}}

\newcommand{\mbh}{M_{\bullet}}
\newcommand{\md}{M_{\rm{}d}}
\newcommand{\sun}{\odot}

\begin{document}

\title{Interaction between stars and an inactive accretion disc\\
 in a galactic core}

\author{Vladim\'{\i}r Karas}
\affiliation{Astronomical Institute, Academy of Sciences, Prague, Czech Republic}
\affiliation{Charles University, Faculty of Mathematics and Physics, Prague, Czech Republic}

\author{Ladislav \v{S}ubr}
\affiliation{Charles University, Faculty of Mathematics and Physics, Prague, Czech Republic}

\begin{abstract}
We discuss the structure of a central stellar cluster whose dynamics is
influenced by gravitation of a supermassive black hole and by the
dissipative interaction of orbiting stars with an accretion disc. We
also take the effect of disc self-gravity into account. We show that the
cluster properties are determined predominantly by the radial profile of
the disc surface density. To this aim we develop a simple steady-state
model of the central cluster and we estimate the rate at which stars
migrate to the centre.

This model is relevant for central regions of an active galactic nucleus
(AGN) containing a rather dense accretion medium. In passing we also 
briefly mention a possibility that a fossil (inactive) disc could
exist in the centre of our own Galaxy. Such a hypothetical disc could
perturb the motion of stars and set them on highly elliptic trajectories
with small pericentre distances. The required mass of the disc is less
than one percent of the central black hole mass, i.e.\ below the upper
limit permitted by present accuracy of the orbital parameters of S-stars
in the Galactic Centre. 
\end{abstract}
\maketitle

\thispagestyle{fancy}

\section{Introduction}
We examined the long-term orbital evolution of stars forming a dense
stellar cluster surrounding a central black hole with an embedded
accretion disc.  This configuration is relevant for central regions of
active galactic nuclei \cite{rauch95,syer91} and it may be
applied also to the center of our own Galaxy,  assuming that rapid
accretion took place and a gaseous disc was formed at some stage of its
history \cite{nayakshin04}. Indeed, the presence of a dissipative
gaseous environment can provide a mechanism driving stars towards the
black hole, while the gravitational influence of the disc may pump
eccentricity of the orbits to large values at some moments of the
evolution \cite{subr04,subr05}. This would help us to understand the
puzzling nature of young stars in the close vicinity of the black hole
in the centre of our Galaxy. An attractive feature of this calculation
is that it provides a well-defined model allowing to estimate the expected
time-scales of the orbital migration as well as the distribution of
eccentricities. It may turn out to be more likely that a different
(non-standard) kind of a gaseous disc or a dusty torus plays the
role, but this would not change the essence of the model.

\begin{figure*}[tb]
\includegraphics[width=\textwidth]{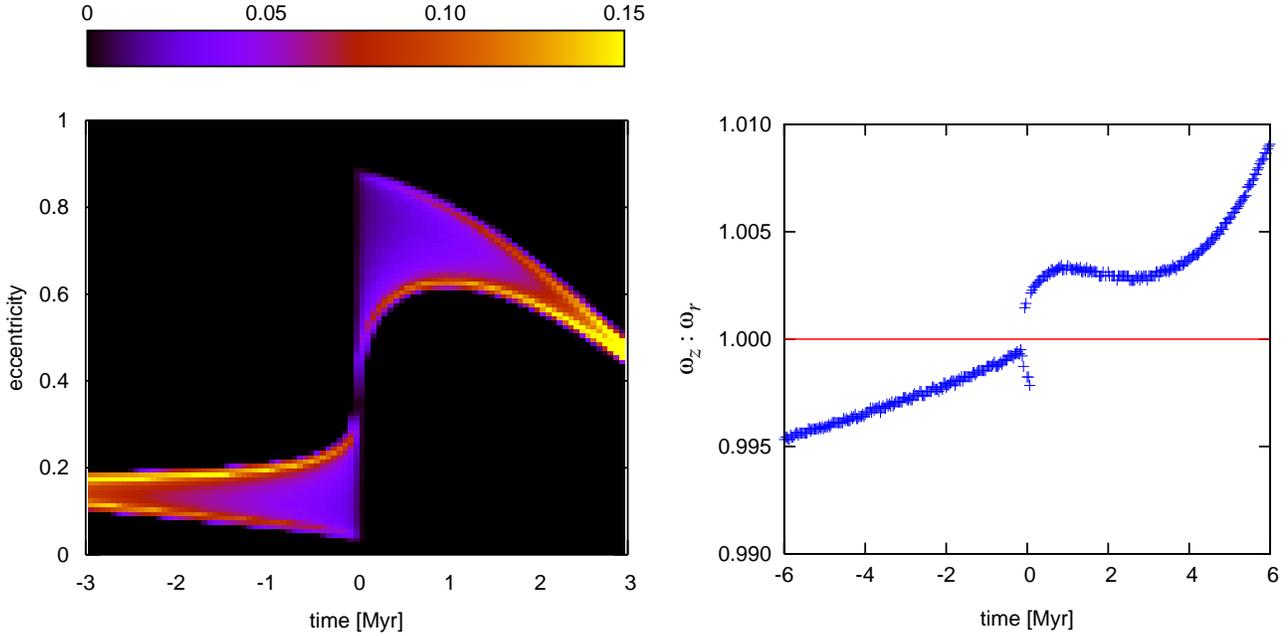}
\caption{An example of the episode during which the stellar orbit becomes
highly elliptical. Left: occasional jumps toward high eccentricity occur 
due to the Kozai-type mechanism in the field of the
central black hole and a gravitating disc. Prevailing circularisation is
expected because of continuing energy losses and the resulting orbital
decay via star-disc collisions, however, fraction of highly eccentric
orbits always persists in the cluster, as indicated by the colour scale. 
These trajectories bring stars close to the
black hole where they can be preferentially captured or destroyed.
Right: the ratio of mean latitudinal frequency $\omega_z$ to the
mean radial epicyclic frequency $\omega_r$ (for the same 
orbit as on the left). It turns out that the 
jump of orbital parameters occurs at the moment when the ratio
$\omega_z/\omega_r=1$. The indicated time-scale corresponds to 
the case of $\mbh=4\times10^6M_{\odot}$ and a hypothetical disc
with the mass $\md=0.01\mbh$.}
\label{fig1}
\end{figure*}

\section{Results}
We idealise a galactic core as a system consisting of a central black
hole, an accretion disc and a  dense stellar cluster. The three
components interact with each other. Naturally, various approximations
need to be imposed in order to keep our model sufficiently simple and
tractable. The aim is to examine the structure of a stellar system in
the region of black hole gravitational dominance $R_{\rm h}$, including
the effects due to a gaseous disc. Two regions of the cluster can be
distinguished according to the characteristic time-scale of processes
dominating the stellar motion. The outer cluster is assumed to reach a
gravitationally relaxed form \cite{bahcall76}, acting as a reservoir of
fresh satellites that are being continuously injected inwards. The
inner cluster is defined as a region where star-disc collisions take
over.

\subsection*{Individual orbits}
Stars loose their orbital energy and momentum by means of successive
dissipative passages through an accretion disc. If the orbit is
inclined with respect to the disc plane, each encounter with
the disc slab slightly modifies the orbital parameters.
The overall trend is to circularize orbits and to decline them 
into the plane of the disc \cite{syer91}.
Characteristic time of aligning stellar orbits with the disc is
$t_{\rm d}(a) \approx t_0 \, M_8
\left( {\Sigma_\ast}/{\Sigma_\odot}\right)^{-1}
\left( {a}/{R_{\rm g}} \right)^{q}\, {\rm yr} \,,$
where $a$ is semi-major axis, $t_0$ and $q$ are constants (order of unity
for the standard disc model) and $M_8\equiv\mbh/(10^8M_\odot)$ is the
fiducial value of the central mass in AGN \cite{karas01}.

Different modes of radial migration apply to stars in the disc plane,
depending whether a star succeeds to open a gap in the disc medium, or
if it remains embedded entirely and proceeds via density waves
excitation. We switch between relevant modes in the numerical
integration.

\begin{figure*}[tb]
\includegraphics[width=\textwidth]{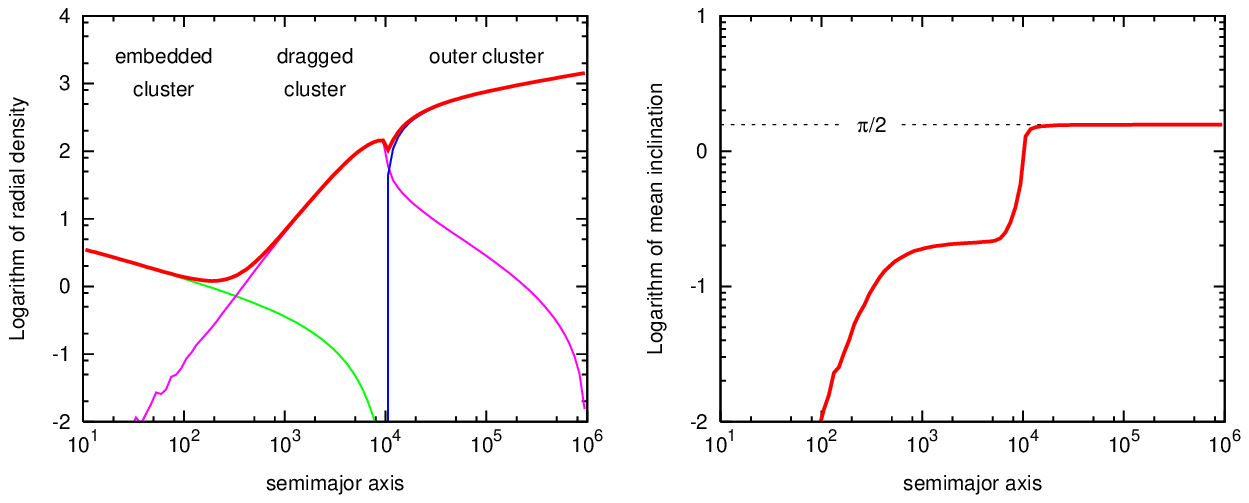}
\caption{Left panel: distribution of
semi-major axes throughout the stellar cluster, modified by interaction
with the gas-pressure dominated standard accretion disc. The broken
power-law profile is established by different types of interaction
governing different regions of phase space. The three stellar
sub-samples are present in our model and they can be easily
distinguished in the plot: the outer reservoir (magenta), the inner part
of the cluster which is dragged by the disc (blue), and a subsample of
stars fully embedded in the disc (green). Right panel: Graph of mean
inclination $\langle{i}\rangle$ in the cluster. The reservoir is
spherically symmetric ($\langle{i}\rangle=\pi/2$), the dragged cluster
is somewhat flattened ($0<\langle{i}\rangle<\pi/2$), and the embedded
population is located in the disc plane ($\langle{i}\rangle=0$).}
\label{fig2}
\end{figure*}

\begin{figure*}[tb]
\includegraphics[width=\textwidth]{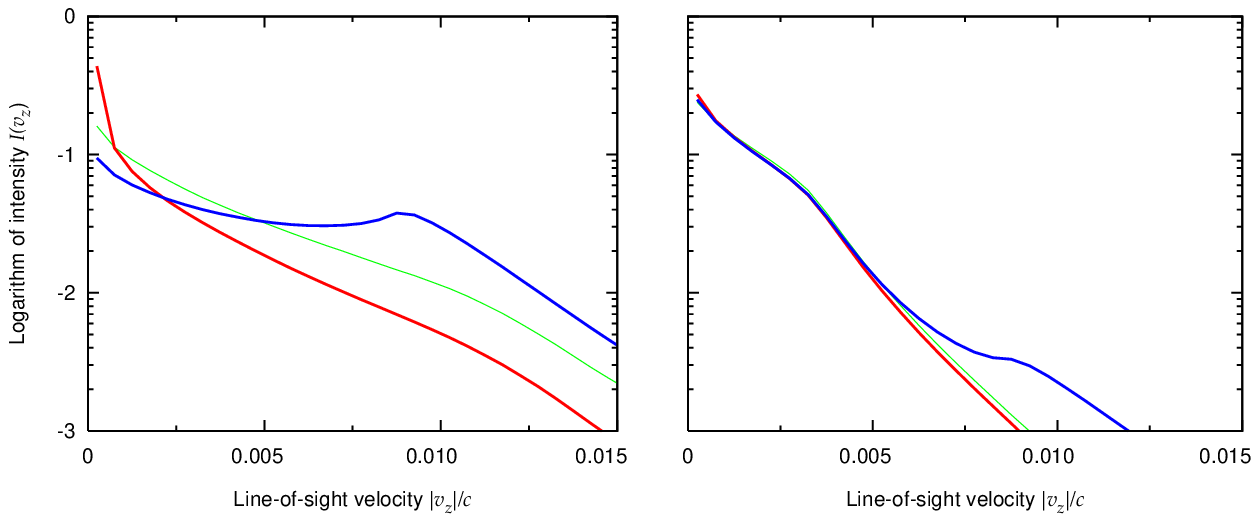}
\caption{Velocity profile along the line of sight
of the inner cluster, integrated across a column of cross-sectional
radius $R_{\rm{}s}=10^4 R_{\rm{}g}$ (left panel) and $R_{\rm{}s}=10^5
R_{\rm{}g}$ (right panel). Blue/red lines represent different view
angles of the observer: $\xi=0^\circ$ and $60^\circ$, respectively.
Growing anisotropy of the modified cluster produces the dependence of
measured line profile on $\xi$, i.e.\ $I{\equiv}I(v_z;\xi)$. The green
line is for the reference Bahcall--Wolf distribution, which
exhibits spherical symmetry. See ref.\ \cite{subr04} for further
details.}
\label{fig3}
\end{figure*}

A gap is cleared in the disc if the embedded star is sufficiently
massive and its Roche radius exceeds the characteristic vertical 
thickness of the disc slab at a corresponding radius,
$r_{_{\rm{}L}}\approx(M_\ast/M)^{1/3}r\,\gta\,h(R)$. 
In this case the motion of the star is coupled with the disc inflow,
\begin{equation}
 \dot{a}_{\rm{}gap}=f(a,\mu_{\rm E},\mbh;b,q_i)
\end{equation}
with
$$
 f(a,\mu_{\rm E},\mbh;b,q_i)\equiv-bM_8^{q_1}\mu_{\rm E}^{q_2}\!
 \alpha^{q_3}
  \left(\frac{a}{R_{\rm g}}\right)^{-q_4}
$$
with $\mu_{\rm E}$ being the accretion rate in 
units of the Eddington accretion rate and $\alpha$ viscosity parameter.
The numerical factor $b$ and power-law indices $q_{1\ldots4}$ are 
determined by
details of the particular model adopted for the disc \cite{karas01}.
  
On the other hand, if the star is too tiny to create the gap, than 
density-waves are the dominating migration process. The resulting
orbital decay can be written in the form
\begin{equation}
 \dot{a}_{\rm{}dw}=\frac{M_\ast}{M_{\sun}}\,f(a,\mu_{\rm E},\mbh;b^\prime,q_i^\prime).
\label{dadtdw}
\end{equation}
Differences in the rate of stellar migration are thus introduced
in the model already within this very simplified picture where the
process of satellite sinking is driven by the gas medium. The dependence
on the orbital parameters, the stellar masses and sizes of stars
causes the gradual segregation of different stellar types in the 
nuclear cluster.

The gravitational field of the disc provides a perturbation
capable of exciting large fluctuations of eccentricity.
Assuming that these fluctuations occur on the time-scale substantially
longer than the orbital period, one can apply an averaging techique
and evolve equation for mean orbital parameters.
See e.g.\ Kozai \cite{kozai62} and Lidov \& Ziglin \cite{lidov76}
for a general introduction to the formalism; see Vokrouhlick\'y \&
Karas \cite{vokrouhlicky98} and \v{S}ubr \& Karas \cite{subr05} 
for its modification to the present situation. 
We computed the gravitational
field of the disc and took its effect into account for the long-term
orbital decay of stellar trajectories. The effect of eccentricity
oscillations is shown in Figure~\ref{fig1}. Here, we have set the
current (time zero) values of orbital eccentricity $e(t)$ and semimajor
axis $a(t)$ identical as those reported for S2 star in
Sagittarius~A$^{\star}$. For the central mass we adopted
$\mbh=4\times10^6M_{\odot}$. A false-colour plot shows what fraction of
time the orbiting star spends with a certain value of the orbital
eccentricity. Therefore, this graph illustrates the expected
fluctuations of eccentricity and explains the existence of highly
elliptic trajectories, assuming that an accretion disc or a torus was
present at the centre of our Galaxy. The required mass of the disc $\md$
was of the order of fraction of percent of the central black hole mass,
consistent with the present-day upper limit. Characteristic time-scale
$t_\mathrm{c}$ of the oscillations depends on $\md/\mbh$ ratio, and is
indeed much longer that the dynamical time at corresponding radius
(of the order of thousand orbital revolutions in our case,
$\md/\mbh\lta0.01$).

\begin{figure*}[tb]
\includegraphics[width=\textwidth]{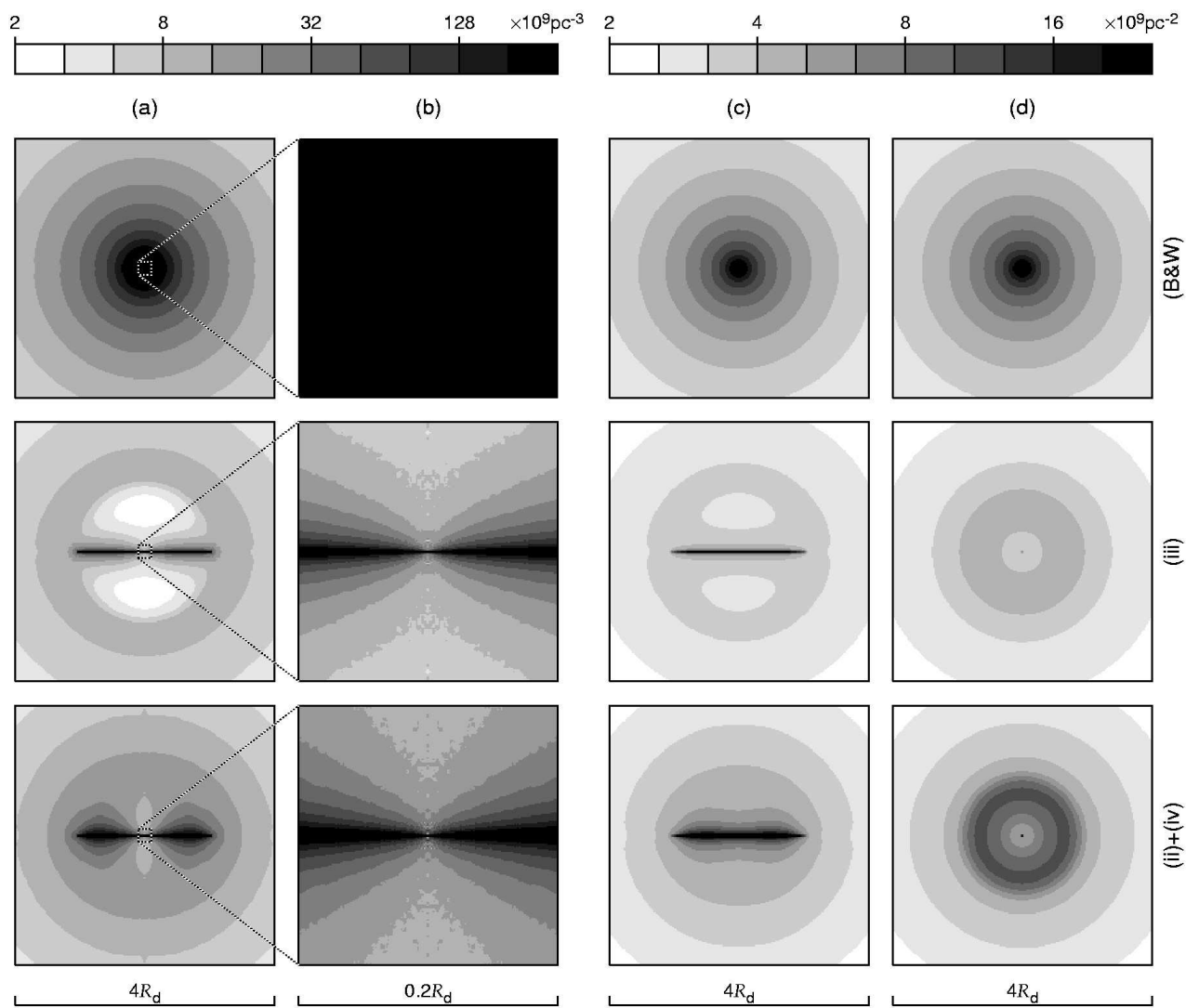}
\caption{The spatial density $n_\ast(r)$ and the corresponding projected
density of the cluster are shown using logarithmically spaced levels of
shading. Columns {\sf{}(a)} and {\sf{}(b)} represent the meridional
section at two different scales, namely, $4R_{\rm d}$ and $0.2R_{\rm d}$
across (radii are expressed in terms of the disc outer radius, $R_{\rm
d}$). Next columns are the edge-on {\sf{}(c)} and the face-on {\sf{}(d)}
projections of the cluster. Across columns, the upper row shows the
referential cluster \cite{bahcall76} ($n_{\ast}\,\propto\,r^{-7/4}$). In
subsequent rows, the system has been already modified via the
interaction with two types of discs, case (iii) and case (ii)+(iv), as
discussed in ref.~\cite{subr04}.}
\label{fig4}
\end{figure*}

\subsection*{The cluster}
Now we construct a steady-state cluster assuming that it is supplied
with fresh stars from the reservoir at a rate inversely proportional to
the relaxation time. The effects introduced in the previous section
are taken into account. Our computational scheme allows us to further
distinguish between two subsamples of the inner cluster: the dragged
inner cluster consists of stars on orbits crossing the disc
periodically; the embedded inner cluster is formed by stars entirely
aligned with the disc. See \v{S}ubr et al.\ \cite{subr04} for details of
our approach. For the definiteness of examples we assumed that the
central mass ($\mbh=10^8M_\odot$) is surrounded by the gas-pressure
dominated Shakura-Sunyaev disc ($s=-3/4$) with $\dot{M}=0.1
M_{\rm{}Edd}$ and viscosity parameter $\alpha=0.1$ (so these values
are adequate for an AGN). The outer stellar
cluster can be characterized by the number density $n_0=10^6 {\rm pc}^{-3}$
and velocity dispersion $\sigma_{\rm{}c} = 200\, {\rm km/s}$.

\subsection*{Structure of the cluster modified by the interaction 
with an accretion disc}
Figure~\ref{fig2} shows the density structure of the modified cluster.
Majority of stars forming the embedded cluster sink to the centre in the
regime of density waves, hence $v_r\propto r^{1/2}$ and $n(a) \propto
a^{-1/2}$. In the dragged cluster, the orbital decay leads to the
governing index given by $s-1/2=-5/4$, and the corresponding number
density $n(a)\propto a^{5/4}$. The asymptotic profile $\propto a^{1/4}$
of the outer cluster is determined by the initial distribution. Isotropy
of the outer cluster is violated in the inner regions where the mean
inclination saturates at $\approx 0.2$ (dragged cluster).  The influence
of the disc manifests itself in different characteristics of the inner
cluster. For example, dependence of the drag on size and mass of stars
causes gradual segregation of different stellar types present in the
cluster.

Figure~\ref{fig3} shows the integrated  properties of the cluster that
can be compared with observation. We plot the shape of a synthetic
spectral line $I(v_z)$, i.e.\ intensity in the line as a function of
line-of-sight velocity $v_z$ near the projected center of the cluster.
Local maximum of the line occurs around
$v\sim\sin(\xi)v_{\rm{}K}(R_{\rm{}d})$. For some values of model
parameters,  this secondary peak exceeds the central maximum and
dominates the predicted  profile. High-velocity tails of the line
profiles are also noticeably affected in comparison with an unperturbed
form of the outer cluster \cite{subr04}.

\subsection*{Oblateness of the cluster}
Figure~\ref{fig4} shows various two-dimensional sections of the
cluster arranged in four columns. One can clearly observe the impact
that star--disc collisions have on the cluster structure, namely, an
increasing oblateness of the stellar population in the core and, in some
cases, the tendency to form an annulus of stars. The reason for
different structures is the continuous crashing of stars on the disc
plane. Furthermore, in case of stars embedded in the disc, different
modes of star-disc interaction occur and facilitate their radial
transport to the central hole.

It is worth noticing that the gravity of the oblate cluster adds with
the field of the embedded accretion disc. This way the effect of orbital
oscillations, discussed in previous sections, may be further enhanced.
Indeed, any non-spherical perturbation, for example general-relativity
effects of frame-dragging near a rotating black hole, can potentially be
responsible for similar kind of oscillations. However, the issue cannot
be resolved within the framework of our simple model, because other
effects intervene (for example the fragmentation of the disc
into selfgravitating clumps) and these may act in the opposite direction.

\section{Discussion}
The orbits are aligned and circularized at typical radii of
$10^2\div10^3 R_{\rm g}$, spiralling further on nearly circular orbits
towards the centre. This provides limits on gravitational waves emerging
from the cluster. The rate of the capture events can be estimated as
\begin{eqnarray}
\dot{M}_{\rm s} & \approx & 10^{-2} M_8^{5/4} \,
\left( \frac{n_0}{10^6\, {\rm pc}^{-3}} \right)^{\!2}
 \nonumber \\
 && \times\left( \frac{M_\ast}{M_\odot} \right)^{\!2}
\left( \frac{R_{\rm d}}{10^4 R_{\rm g}} \right)
M_\odot\,{\rm yr}^{-1}\,. 
\end{eqnarray} 
The total accretion rate onto the black hole is a sum of $\dot{M}_{\rm
s}$ (which involves massive stars in the cluster), and the accretion
rate $\dot{M}$ (gas in the disc). 

It is worth noticing that the orbital decay of stars near a black hole
is indeed relevant for forthcoming gravitational wave experiments,
because the gas-dynamical drag should be taken into account with
sufficient accuracy.  It was estimated \cite{glampedakis02,narayan00}
that this effect can be safely ignored at late stages, shortly before
the star plunges into the hole, if accretion takes place in the mode of
a very diluted flow, but  the situation is quite different in case of
AGN hosting rather dense nuclear discs. Gas-dynamical effects most
likely dominate over the gravitational radiation as far as their
influence on the cluster structure is concerned.

The average rate of energy losses which the orbiter experiences
via gravitational radiation (per revolution) can be written in terms
of orbital parameters \cite{peters63}:
\begin{equation}
 \dot{\cal E}=\frac{32}{5}\frac{G^4}{c^5}
  \frac{M^3M_{\ast}^2}{a^5\left(1-e^2\right)^{7/2}}\;
 f_1(e).
 \label{degw}
\end{equation}
Linked with this $\dot{\cal E}$ is the change of semi-major axis
\begin{equation}
  \dot{a} = -1.28\times10^{-7}c
   M_8^{-1}\frac{M_{\ast}}{M_{\sun}}
    \left[\frac{a}{r_{\rm g}}\right]^{-3}\;
 f_2(e)
\end{equation}
and the loss of angular momentum
\begin{equation}
   \dot{\cal L}=
    \frac{32}{5}\frac{G^{7/2}}{c^5}
    \frac{M^{5/2}M_\ast^2}{a^{7/2}\left(1-e^2\right)^{2}}\;
 f_3(e).
 \label{dlgw}
\end{equation}
Functions $f_1$, $f_2$ and $f_3$ are of the order of unity. 
The above-given formulae (\ref{degw})--(\ref{dlgw}) assume the 
star is on an elliptic trajectory in Schwarzschild geometry.
It therefore provides only an order-of-magnitude estimation, because
collective effects operating in the cluster are neglected.

Gravitational radiation orbital decay competes with the orbital decay 
caused by hydrodynamics of star--disc encounters. The relative 
importance of these two influences depends on the ratio between
the hydrodynamical drag (which is roughly proportional to the disc
density) versus gravitational-wave losses (which increase with 
orbital eccentricity). Discussing this interplay is a tricky task
once the gravitationally induced oscillations of eccentricity
are taken into account. See refs.\ \cite{blaes02,karas01} for a 
more detailed discussion of these effects. Substantially different
results and time-scales can be expected for objects with rather 
dense discs (such as the case of standard discs in AGNs) and
those which are inactive and represent a rather clean system 
(such as our own Galactic Center).

\section{Conclusions}
Although the above-described model is
intended mainly for active galactic nuclei with a relatively dense
accretion disc or a dusty torus, the problem of stars crashing on a
gaseous disc may be relevant also for the centre of our Galaxy. The
modified cluster structure is relevant for estimating the rate of
black-hole feeding and, vice-versa, for the issue of feedback that a
super-massive black hole exhibits on the the host galaxy. 

The inherent limitations persist in our present discussion, namely, we have
not incorporated a fully self-consistent treatment of the disc gravity.
Even though we computed the gravitational field across a sufficiently
large domain of space, we did not account for the feedback which
star--disc collisions exert on the disc structure. Introducing some kind
of a clumpy model of the disc will be very interesting, as it may exert a
more substantial effect on the cluster structure by elevating the impact
of star--disc collisions. For further details we refer the
reader to papers \cite{subr04} and \cite{subr05}, and to references
cited therein.

The authors gratefully acknowledge support from the Czech Science Foundation
grant No.\ 205/03/0902 and from Charles University in Prague (299/2004).


\bigskip

\begin{thebibliography}{99}
\bibitem{bahcall76}Bahcall J.~N., Wolf R.~A., 1976, ApJ, 209, 214
\bibitem{blaes02}Blaes~O., Lee M.~H., Socrates~A., 2002, ApJ, 578, 775
\bibitem{glampedakis02}Glampedakis K., Kennefick D., 2002, Phys.\ Rev. D, 66, 044002
\bibitem{karas01}Karas V., \v{S}ubr L., 2001, A\&A, 376, 687
\bibitem{kozai62}Kozai~Y., 1962, AJ, 67, 591
\bibitem{lidov76}Lidov M.~L., Ziglin S.~L., 1976, Celest. Mech., 13, 471
\bibitem{narayan00}Narayan R., 2000, ApJ, 536, 663
\bibitem{nayakshin04}Nayakshin~S., Cuadra~J., Sunyaev~R., 2004, A\&A, 413, 173
\bibitem{peters63}Peters P.\,C., Mathews J., 1963, Phys.\ Rev., 131, 435
\bibitem{rauch95}Rauch K.~P., 1995, MNRAS, 275, 628
\bibitem{subr04}\v{S}ubr L., Karas V., Hur\'e J.-M., 2004, MNRAS, 354, 1177
\bibitem{subr05}\v{S}ubr L., Karas V., 2005, A\&A, in press
\bibitem{syer91}Syer~D., Clarke C.~J., Rees M.~J., 1991, MNRAS, 250, 505
\bibitem{vokrouhlicky98}Vokrouhlick\'y~D., Karas~V., 1998, MNRAS, 298, 53
\end{thebibliography}
\end{document}